\documentclass[amsmath,amssymb,
aps, prb, reprint,
superscriptaddress,
]{revtex4-2}
\usepackage[title]{appendix}
\usepackage{graphicx}% Include figure files
\usepackage{dcolumn}% Align table columns on decimal point
\usepackage{bm}% bold math
\usepackage[dvipsnames]{xcolor}
\usepackage{gensymb}
\usepackage{float}
\usepackage{multirow}
\usepackage{textcomp}
\usepackage{gensymb}
\usepackage{hyperref}

\begin{document}

\preprint{APS/123-QED}

\title{Robust phase correction techniques for terahertz time-domain reflection spectroscopy}
\author{Kasturie D. Jatkar}
\affiliation{Department of Physics, Stockholm University, 106 91 Stockholm, Sweden.}
\author{Tien-Tien Yeh}
\affiliation{Department of Physics, Stockholm University, 106 91 Stockholm, Sweden.}
\author{Matteo Pancaldi}
\affiliation{Department of Molecular Sciences and Nanosystems, Ca’ Foscari University of Venice, 30172 Venice, Italy}
\author{Stefano Bonetti}
\email{stefano.bonetti@fysik.su.se}
\affiliation{Department of Physics, Stockholm University, 106 91 Stockholm, Sweden.}
\affiliation{Department of Molecular Sciences and Nanosystems, Ca’ Foscari University of Venice, 30172 Venice, Italy}

\begin{abstract}
% "A Novel Technique for Precise Phase Correction and Optical Property Extraction in Terahertz Reflection Spectroscopy"

We introduce a systematic approach that enables two robust methods for performing terahertz time-domain spectroscopy in reflection geometry. Using the Kramers-Kronig relations in connection to accurate experimental measurements of the amplitude of the terahertz electric field, we show how the correct phase of the same field can be retrieved, even in the case of partly misaligned measurements. Our technique allows to accurately estimate the optical properties of in principle any material that reflects terahertz radiation. We demonstrate the accuracy of our approach by extracting the complex refractive index of InSb, a material with a strong plasma resonance in the low-terahertz range. Our technique applies to arbitrary incidence angles and polarization states.

\end{abstract}

\maketitle

\section{\label{sec:intro}Introduction}
 
Terahertz (THz) spectroscopy is a well-established, non-destructive measurement technique which has demonstrated its usefulness in various fields such as solid state physics \cite{cmp_Lee_2009}, chemistry \cite{chemical_Fischer_Helm_Jepsen_2007, review_Baxter_Guglietta_2011}, biology \cite{bio_Globus_Woolard_Khromova_Crowe_Bykhovskaia_Gelmont_Hesler_Samuels_2003, bio_Son_2009}, pharmaceuticals \cite{pharma_McGoverin_Rades_Gordon_2008}, security \cite{security_Palka_Szustakowski_Kowalski_Trzcinski_Ryniec_Piszczek_Ciurapinski_Zyczkowski_Zagrajek_Wrobel_2012}, and imaging \cite{imaging_Jepsen_Cooke_Koch_2011}, 
to cite a few. Since THz radiation lies between the microwave and infrared regions, its energy range from 0.4 to 40 meV is uniquely suited for studying low-energy excitations such as phonons, magnons, polarons \cite{phonons_magnons_Saln_2019, emerging_Spies_Neu_Tayvah_Capobianco_Pattengale_Ostresh_Schmuttenmaer_2020, polaron_Zheng_2021}. THz has also been considered as a viable candidate for detecting axions \cite{schutte-engel2021axiona,dona2022design}. The process of studying material properties using THz spectroscopy has been improving over time with the advancements in the generation and detection of THz radiation, both in terms of frequency range and signal-to-noise ratio (SNR). One of the most widely used techniques enabled by this development is the  THz time-domain spectroscopy (THz-TDS). In a THz-TDS measurement, the acquired data contains information about the amplitude as well as phase of the electric field, which allows the study of complex, frequency-dependent optical properties of materials \cite{Withayachumnankul_Naftaly_2014, Jepsen_2019}. 

THz-TDS is often performed in transmission geometry since the process of obtaining optical properties in this configuration is relatively straightforward. This arrangement is mainly useful for measuring transparent materials such as polymers \cite{polymers_Jansen_Wietzke_Koch_2013}, which have low absorption in the THz range. Samples that show strong absorption are usually measured by attenuated total reflection spectroscopy \cite{ATR_Hirori_Yamashita_Nagai_Tanaka_2004}. Samples characterized by low THz transmission, the majority of known materials, require measurements in reflection geometry. Such a geometry, in fact, allows for exploring even metallic and superconducting samples, which screen THz radiation very efficiently. However, measurements of the phase of the electric field in reflection geometry are extremely sensitive to the relative alignment of sample and reference \cite{pump_phase_Pashkin_2003, Si_Nashima_Morikawa_Takata_Hangyo_2001}. Inaccurate positioning produces large errors in the results, increasing the complexity of post-processing for data extraction. To this end, several computational techniques have been developed, such as the maximum entropy method (MEM) \cite{mem_Vartiainen, mem2_Unuma}, singly subtractive Kramers-Kronig (SSKK) \cite{sskk_Lucarini_Ino_Peiponen_Kuwata-Gonokami_2005}, multiply subtractive Kramers-Kronig (MSKK) \cite{mskk_Palmer_Williams_Budde_1998}, differential multiply subtractive Kramers-Kronig (DMSKK) \cite{dmskk_Granot_Ben-Aderet_Sternklar_2008}. These techniques require heavy iterations or the knowledge of so-called ``anchor points'' (i.e., points where the properties of the samples are known) to converge to accurate values. To circumvent the challenges created due to phase sensitivity in reflection geometry, several innovative experimental techniques were proposed, as exemplified by Ref.~\cite{pump_phase_Pashkin_2003, through_sub_Lin_Burton_Engelbrecht_Tybussek_Fischer_Hofmann_2020,Si_Nashima_Morikawa_Takata_Hangyo_2001, Howells_Schlie_1996}. Others have also attempted to combine results from different setups to verify their findings, as seen in Ref.~\cite{TR_ref_Khazan_Meissner_Wilke_2001, STO_tr_Mori_Igawa_Kojima_2014, Linbo3_Igawa_Mori_Kojima_2014}. Despite their effectiveness, these techniques often tend to be material-specific and cannot be universally applied to all sample types. This has limited their widespread adoption in the scientific community.

In this work, we introduce a systematic approach that greatly simplifies the process of extraction of information from THz-TDS in reflection geometry. Our technique, based on the Kramers-Kronig relations, can retrieve the correct phase value from in principle any arbitrary shift between the sample and reference. We provide a detailed description of the experimental setup and the mathematical derivation of our approach, which is implemented in two ways. The first method follows a relatively straightforward fitting to an analytical function that is derived and presented here. The second method uses an iterative approach which can be either simulated or performed experimentally to find the shift. In the companion paper Ref.~\cite{UDCM} we show the applicability of our approach to the cases of Si and InSb samples, measured at normal incidence. Here, besides the more detailed derivation, we extend the validation of the method to arbitrary incidence angle and polarization direction for the case of InSb.

\begin{figure*}[t]
\includegraphics[width=\textwidth]{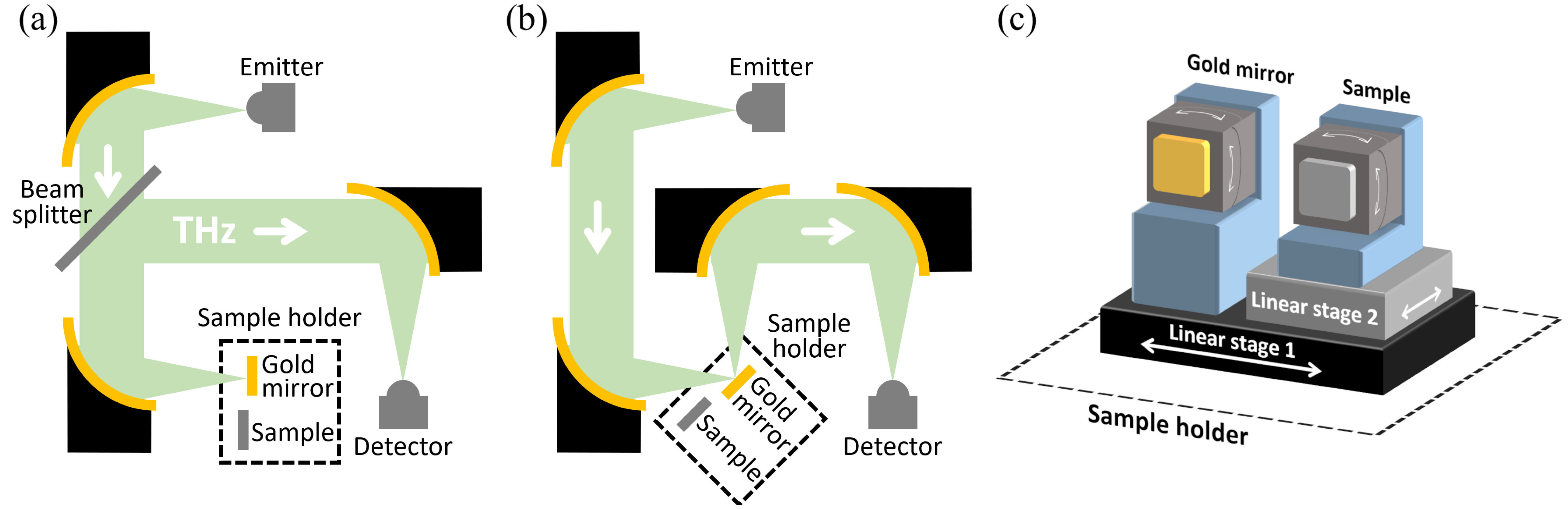}
\caption{\label{fig:1} Schematic of the THz-TDS setup in reflection geometry for (a) normal and (b) 45$\degree$ incidence of the THz radiation. The detector and the emitter are based on photoconductive antennas. Beam splitter in (a): intrinsic Si wafer. (c) Detailed schematic of the sample holder. Both the reference (gold mirror) and the sample stages include a double goniometer. The first linear stage is a standard stepper motor, the second one is a piezoelectric stage.}
\end{figure*}%

\section{\label{sec:exp}Experimental Setup}

%\subsection{\label{sec:expsetup}Experimental setup}
The experimental THz-TDS setup is shown schematically in Fig.~\ref{fig:1}. THz radiation is generated and detected using an integrated laser system based on photoconductive antennas (TeraFlash Pro, TOPTICA Photonics Inc.). The setup has a time resolution of 0.05 ps, and a variable scanning range that we set to 70 ps. The setup is arranged in two different ways to investigate the effects of the angle of incidence of the THz radiation on the sample. For normal incidence geometry, a 3 mm thick intrinsic Si beam splitter is employed to collect the reflected THz beam from the sample/reference as shown in Fig.~\ref{fig:1}(a). For 45\degree -incidence geometry, an additional off-axis parabolic mirror is included to collect the reflected THz beam as illustrated in Fig.~\ref{fig:1}(b). At the focal point, our experimental setup features a custom-made sample stage, depicted in Fig.~\ref{fig:1}(c). This stage is constructed using two linear stages and two goniometers, providing precise control over the relevant degrees of freedom as tilt correction plays an essential role in the proper implementation of this technique. The sample and reference (a thick gold mirror) are each mounted on separate stages. Additionally, the sample is placed on a piezoelectric stage, enabling precise positioning in the direction perpendicular to the sample surface. This entire assembly is then placed on a motor-controlled stage that translates parallel to the sample surface. This stage helps to minimize the impact of humidity-induced drift in the collected data, by implementing a procedure of cycling between the sample and reference for each measurement, conducted at 10-second intervals. In 10 seconds, up to 100 averages are obtained, and this process is repeated to improve the SNR. %Other important considerations taken into account for the experimental optimization and to mitigate the effects of humidity are detailed in Appendix \ref{app:A} and \ref{app:B}.

\section{\label{sec:prm}Theory}
\subsection{\label{sec:fundamentals_PRM}Fundamentals of the phase correction technique}

In a typical THz-TDS measurement in reflection geometry, the THz pulses reflected from the sample and reference are recorded in the time domain. The reflection coefficient is determined by calculating the ratio of the fast Fourier transform (FFT) of the time traces measured from the sample and, respectively, the reference. The resultant measured complex reflection coefficient $\tilde{r}_m(\omega)$ as a function of frequency $\omega$ is written as
\begin{equation}
\label{eq:rm}
\tilde{r}_m(\omega)=\left|\frac{\widetilde{E}_{s}(\omega)}{\widetilde{E}_{r}(\omega)}\right| e^{-i\varphi_m}  = |\tilde{r}_m(\omega)|e^{-i\varphi_m},
\end{equation}
where $\widetilde{E}_{s}$, $\widetilde{E}_{r}$ represent the complex electric field reflected off the sample and, respectively, off the reference. The measured relative phase given by $\varphi_m(\omega)$ is sensitive to the sample positioning along the light propagation direction. Generally, after the relative tilt between sample and reference has been experimentally corrected, the measured phase can be expressed as
\begin{equation}\label{eq:phi_m}
\varphi_{m}(\omega)=\varphi_{i}(\omega)+\frac{\omega}{c}\frac{2l}{\cos{\theta}},
\end{equation}
where $\varphi_i$ is the intrinsic phase induced by the sample and $2l$ denotes the excess traveling distance of the THz pulse, resulting from an unknown shift $l$ in the sample position with respect to the reference. The angle of incidence of the THz pulse is taken into account with the $\cos{\theta}$ factor.

\begin{figure}[t]
\includegraphics[width=\columnwidth]{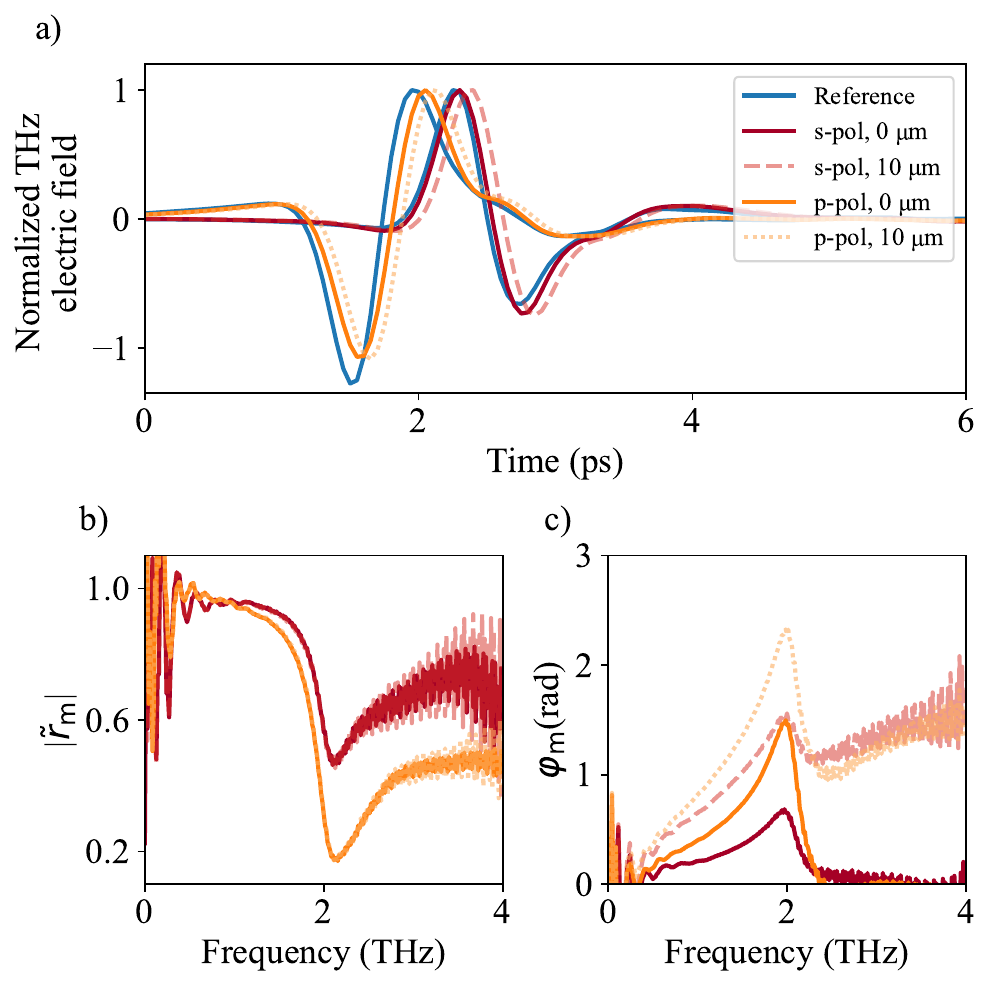}
\caption{\label{fig:2} (a) Time traces from 45\degree-incidence THz reflection measurements with a gold mirror as reference (blue) and InSb as sample for two different positions (labeled in legend) with $s$- and $p$-polarized THz radiation. (b) Corresponding FFT amplitudes and (c) FFT phases of the reflection coefficient of InSb at two different positions for each polarization.}
\end{figure}

Figure \ref{fig:2}(a) presents typical experimental outcomes obtained from a THz-TDS measurement. Here we show the time traces acquired from the measurement of the sample (in this case, single crystal InSb) and reference (gold), with the THz radiation incident at an angle of 45\degree ~for $s$- and $p$-polarized fields. The inversion of the time trace for the two orthogonal polarizations is consistent with the Fresnel coefficients in the two cases. The measurements are performed at two positions of the piezoelectric stage with a relative change of 10 \textmu m for the THz optical path. The retrieved amplitude $|\tilde{r}_m|$ and phase $\varphi_m$ of the complex reflection coefficient $\tilde{r}_m$ for these two positions are plotted in Fig.~\ref{fig:2}(b) and (c). As predicted by Eq.~\eqref{eq:phi_m}, the shift in the temporal position of the peaks in Fig.~\ref{fig:2}(a) is reflected in a change of slope for the phase. On the other hand, $|\tilde{r}_m|$ undergoes no major modifications while changing the sample position, apart from the increased noise in the high-frequency region. In the following, me make use both these features, for the phase and the amplitude, to devise a strategy based on the Kramers-Kronig relations to reliably correct the phase and extract the correct optical parameters.

The Kramers-Kronig relations are bidirectional complex relations that work on the principle of causality and are commonly used in optics to connect the real and imaginary parts of a response function, typically the optical permittivity. An alternative formulation has been derived for the complex reflection coefficient $\tilde{r}=|\tilde{r}|e^{-i\varphi}$, exploiting the fact that $\ln(\tilde{r})=\ln|\tilde{r}|-i\varphi$ \cite{smith1977dispersion,lovell1974applicationa, roessler1965kramerskronig, nash1995kramerskronig, peiponen1997ii, bertie1996accurate}. The alternative formulation is written as
\begin{equation}\label{eq:kkr}
   \varphi(\omega) = \varphi_0 +  \frac{2\omega}{\pi}\textbf{P}\int_0^\infty\frac{ \text{ln}|\tilde{r}(\Omega)|}{\Omega^2-\omega^2}d\Omega,
\end{equation}
\begin{multline}
    \label{eq:ikkr}
  \ln\left|\tilde{r}(\omega)\right| - \ln\left|\tilde{r}(\omega^{\prime})\right| = \ln\left|\frac{\tilde{r}(\omega)}{\tilde{r}\left(\omega^{\prime}\right)}\right| =\\
  \frac{2}{\pi} \textbf{P} \int_0^{\infty} \Omega \varphi(\Omega)\left(\frac{1}{\Omega^2-\omega^2}-\frac{1}{\Omega^2-\omega^{\prime 2}}\right) d \Omega,
\end{multline}
where \textbf{P} denotes Cauchy's principal value. These two equations state that the phase of the reflection coefficient can be extracted from its amplitude and, conversely, the amplitude of the reflection coefficient can be extracted from its phase. We refer to Eq.~\eqref{eq:kkr} as the \emph{direct} Kramers-Kronig relation and to Eq.~\eqref{eq:ikkr} as the \emph{inverse} form. The $\varphi_0$ term is mostly known to have a value of zero for insulators and metals \cite{Jepsen_2019, smith1977dispersion,STO_tr_Mori_Igawa_Kojima_2014,Howells_Schlie_1996}. The term $\ln|\tilde{r}(\omega')|$ denotes a known value of $|\tilde{r}|$ at a reference frequency $\omega'$ obtained from a measurement. Since the integral of the phase does not necessarily return a unique value for the reflection coefficient, subtraction of a known value from the extracted curve ensures correct results, as described in detail in Ref.~\cite{smith1977dispersion, peiponen1997ii, nash1995kramerskronig}.

The Kramers-Kronig relations require a large frequency range, nominally the entire range from 0 to infinity. Working with them within the limited accessible frequency range of a typical THz-TDS setup requires some careful considerations and assumptions. We start by separating the integral term in Eq.~\eqref{eq:kkr} in two parts, one within and one outside the measurement range bounded by the frequency $\omega_{\text{end}}$: 
\begin{multline}\label{eq:kkr_split}
     \varphi(\omega)\approx\varphi_0+\frac{2 \omega}{\pi} \mathbf{P} \int_0^{\omega_{\text {end }}} \frac{\ln \left|\tilde{r}\left(\Omega\right)\right|}{{\Omega}^2-\omega^2} d \Omega   \\
    + \frac{2 \omega}{\pi} \mathbf{P} \int_{\omega_{\text {end }}}^{\infty} \frac{\ln \left|\tilde{r}\left(\omega_{\text {end }}\right)\right|}{{\Omega}^2-\omega^2} d \Omega.
\end{multline}

The second integral of Eq.~\eqref{eq:kkr_split} describes the high-frequency term, beyond the experimental measurement range, where we have assumed that the amplitude of the reflection coefficient remains constant with the same value measured at the edge of the measurement range, i.e.\ equal to the value of $|\tilde{r}(\omega_\text{end})|$. This assumption is physically justified as long as the measurement range is large enough to fully contain possible resonances. Our experiments on InSb, shown in Fig.~\ref{fig:2}(b) and in the Letter associated to this work \cite{UDCM}, show that in the case of a Drude-like response with a plasma edge at 2 THz, such assumption is satisfied for $\omega_{\text {end}}>3$ THz. Equation \eqref{eq:kkr_split} can then be rewritten as
\begin{multline}\label{eq:phi_kkr_full}
    \varphi(\omega)\approx\varphi_0+\frac{2 \omega}{\pi} \mathbf{P} \int_0^{\omega_{\text {end }}} \frac{\ln \left|\tilde{r}\left(\Omega\right)\right|}{\Omega^2-\omega^2} d \Omega   \\+\frac{1}{\pi}\ln|\tilde{r}(\omega_{\text{end}})|\ln\left|\frac{\omega_{\text{end}}+\omega}{\omega_{\text{end}}-\omega}\right|.
\end{multline}
All terms in Eq.~\eqref{eq:phi_kkr_full} can in principle be obtained from experimentally measured quantities.  However, in the case of the \emph{direct} relations, the assumption that the reflection coefficient is constant beyond the measured range may be insufficient in the general case, and the limited frequency range available from a THz measurement tends to alter the shape of the calculated phase, which in turn severely affects the results, as shown in detail in Appendix \ref{app:A}.

The goal of this work is to propose the most robust and general method applicable to various materials and experimental configurations. Hence, we choose to focus on the inverse form of the Kramers-Kronig relations shown in Eq.~\eqref{eq:ikkr}, as detailed in the thorough discussion presented in the next paragraphs.

\begin{figure}[t]
    \centering
    \includegraphics[width=0.8\columnwidth]{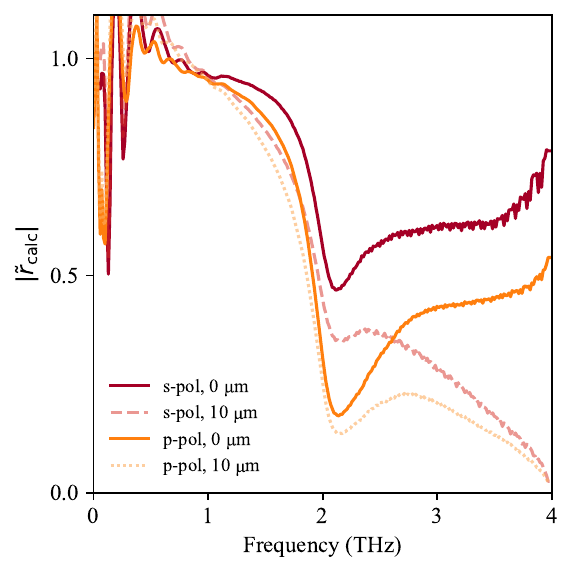}
    \caption{Reflection coefficient amplitude for InSb single crystalline sample calculated by integrating the measured phase (within the 0 - 4 THz range) for $s$- (red) and $p$-polarized (orange) radiation obtained at 0 \textmu m (solid lines) and with a path change of 10 \textmu m denoted by dashed lines.}
    \label{fig:3}
\end{figure}

The main concern in extracting optical properties from the measured phase values is the presence of the extra linear term in $\omega$ due to misplacement, as indicated in Eq.~\eqref{eq:phi_m}. Hence, we need to understand the effect of the misplacement term on the validity of the inverse Kramers-Kronig relation. To this end, we separate the measured phase in its constituents, and the integral in Eq.~\eqref{eq:ikkr} can be expressed as
\begin{multline}
    \label{eq:ikkr_lin_inf}
  \frac{2}{\pi} \textbf{P} \int_0^{\infty} \Omega \varphi_m(\Omega)g(\Omega;\omega, \omega') d \Omega =\\
  \frac{2}{\pi} \textbf{P} \int_0^{\infty} \Omega \varphi_i(\Omega)g(\Omega;\omega, \omega') d \Omega \\
  +\frac{4l}{\pi c\cos{\theta}} \textbf{P} \int_0^{\infty} \Omega^2g(\Omega;\omega, \omega')d \Omega =
  \ln \left| \frac{\tilde{r}_i(\omega)}{\tilde{r}_i(\omega^{\prime})}\right|,
\end{multline}
where, for simplicity, we have defined the difference inside the integral as
\begin{equation}
     g(\Omega;\omega, \omega') \equiv \frac{1}{\Omega^2-\omega^2}-\frac{1}{\Omega^2-\omega^{\prime 2}}.
\end{equation}
The last equality in Eq.~\eqref{eq:ikkr_lin_inf} follows from the fact that the integral of the intrinsic phase from 0 to $\infty$ provides the intrinsic reflection coefficient $|\tilde{r}_{i}|$, and the integral of the misplacement term evaluates to 0, as detailed in the first part of Appendix \ref{app:B}. Since the misplacement only has a minor effect on the amplitude of the complex reflection coefficient, we can safely assume that
$\left|\tilde{r}_i(\omega)\right|\approx\left|\tilde{r}_m(\omega)\right|$. Thus we can write% it turns out that
\begin{equation}
    \ln \left| \frac{\tilde{r}_m(\omega)}{\tilde{r}_m(\omega^{\prime})}\right| = \frac{2}{\pi} \textbf{P} \int_0^{\infty} \Omega \varphi_m(\Omega)g(\Omega;\omega, \omega') d \Omega.
    \label{eq:r_m}
\end{equation}
Eq.~\eqref{eq:r_m} shows that the presence of a misplacement term does not alter the validity of the inverse Kramers-Kronig relation (based on causality) when the integration is ideally performed in the whole frequency range (up to infinity).

However, as discussed before, any real measurement is always bounded by a certain finite $\omega_\text{end}$, which limits the validity of our assumptions. To this end, Fig.~\ref{fig:3} shows the calculated reflection coefficient amplitude $|\tilde{r}_\text{calc}|$, defined according to the following equation, and corresponding to a reduced form of Eq.~\eqref{eq:r_m}:
\begin{multline}
    \label{eq:ikkr_calc}
  \ln\left|\tilde{r}_\text{calc}(\omega)\right| \equiv\\
  \ln\left|\tilde{r}_m(\omega^{\prime})\right| + \frac{2}{\pi} \textbf{P} \int_0^{\omega_\text{end}} \Omega \varphi_m(\Omega)g(\Omega;\omega, \omega') d \Omega,
\end{multline}
where $|\tilde{r}_m|$ and $\varphi_m$ correspond to the measured amplitudes and phases depicted in Figs.~\ref{fig:2}(b) and \ref{fig:2}(c), respectively, $\omega_\text{end} = 4~\text{THz}$ and $\omega' = 1~\text{THz}$. It is evident from Fig.~\ref{fig:3} that the reflection coefficients extracted from the phases with higher slopes (corresponding to a misplacement of 10 \textmu m) exhibit larger deviations. Thanks to Eq.~\eqref{eq:phi_m}, we can precisely evaluate the effect of the misplacement term on the integral in Eq.~\eqref{eq:ikkr_calc}:
\begin{widetext}
\begin{equation}
\begin{aligned}\label{eq:ikkr_expand}
\frac{2}{\pi} \textbf{P} \int_0^{\omega_\text{end}} \Omega &\varphi_m(\Omega)g(\Omega;\omega, \omega') d \Omega =\\
&\frac{2}{\pi} \textbf{P} \int_0^{\omega_\text{end}} \Omega \varphi_i(\Omega)g(\Omega;\omega, \omega') d \Omega + \frac{4l}{\pi c\cos{\theta}} \textbf{P} \int_0^{\omega_\text{end}} \Omega^2 g(\Omega;\omega, \omega') d \Omega =\\
&\ln \left| \frac{\tilde{r}_m(\omega)}{\tilde{r}_m(\omega^{\prime})}\right| - \frac{2}{\pi} \int_{\omega_\text{end}}^{\infty} \Omega \varphi_i(\Omega) g(\Omega;\omega, \omega') d \Omega + \frac{4l}{\pi c\cos{\theta}} \textbf{P} \int_0^{\omega_\text{end}} \Omega^2g(\Omega;\omega, \omega') d \Omega,
\end{aligned}
\end{equation}
\end{widetext}
where we made use of the fact that $\left|\tilde{r}_i(\omega)\right|\approx\left|\tilde{r}_m(\omega)\right|$ and $\omega,\omega'<\omega_\text{end}$. After grouping all the terms directly related to measured quantities on the left-hand side, and the remaining terms on the right-hand side, we obtain
\begin{multline}\label{eq:ikkr_grouped}
    \frac{2}{\pi} \textbf{P} \int_0^{\omega_\text{end}} \Omega \varphi_m(\Omega)g(\Omega;\omega, \omega') d \Omega
    - \ln\left|\frac{\tilde{r}_m(\omega)}{\tilde{r}_m(\omega')}\right| =\\
    \frac{4l}{\pi c\cos{\theta}} \textbf{P} \int_0^{\omega_\text{end}} \Omega^2g(\Omega;\omega, \omega') d \Omega
    + E_{\omega_\text{end}}\left(\omega\right),
\end{multline}
where
\begin{equation}\label{eq:ikkr_error}
    E_{\omega_\text{end}}\left(\omega\right) = -\frac{2}{\pi} \int_{\omega_\text{end}}^{\infty} \Omega \varphi_i(\Omega)g(\Omega;\omega, \omega') d \Omega.
\end{equation}
While Eq.~\eqref{eq:ikkr_grouped} defines how the measured data can be modeled and interpreted in the presence of a sample misplacement, Eq.~\eqref{eq:ikkr_error} expresses the approximation error related to a finite integration range. This approximation error depends on the intrinsic phase $\varphi_i$, which in reflection geometry typically remains very small and deviates from zero only near the presence of a resonance. Therefore, while dealing with the error term involving higher frequencies (above $\omega_\text{end}$), as long as $\omega_\text{end}$ does not lie close to a resonance feature, we can safely assume that the result of the integral is negligible and can be treated as a constant $C \approx E_{\omega_\text{end}}$. Moreover, the integral of the misplacement term in Eq.~\eqref{eq:ikkr_grouped} is finite within the $(0,\omega_{\text{end}})$ range, and is solved for in the second part of Appendix \ref{app:B}. Finally, we get
\begin{multline}\label{eq:ikkr_final}
    \ln\left|\frac{\tilde{r}_\text{calc}(\omega)}{\tilde{r}_m(\omega)}\right| = \\
    \frac{2}{\pi} \textbf{P} \int_0^{\omega_\text{end}} \Omega \varphi_m(\Omega)g(\Omega;\omega, \omega') d \Omega
    - \ln\left|\frac{\tilde{r}_m(\omega)}{\tilde{r}_m(\omega')}\right| \approx\\
    \frac{2l}{\pi c \cos\theta}\left(\omega \ln \left|\frac{\omega_{\mathrm{end}}-\omega}{\omega_{\mathrm{end}}+\omega}\right|- \omega' \ln \left|\frac{\omega_{\mathrm{end}}-\omega'}{\omega_{\mathrm{end}}+\omega'}\right|\right) + C.
\end{multline}
In this form, the difference between $\ln\left|\tilde{r}_\text{calc}(\omega)\right|$, calculated according to Eq.~\eqref{eq:ikkr_calc}, and the measured $\ln\left|\tilde{r}_m(\omega)\right|$ is evaluated in terms of an analytical function explicitly depending on the sample-to-reference misplacement $l$.

\subsection{Analytical fitting method}
\begin{figure}[t]
    \includegraphics[width=\columnwidth]{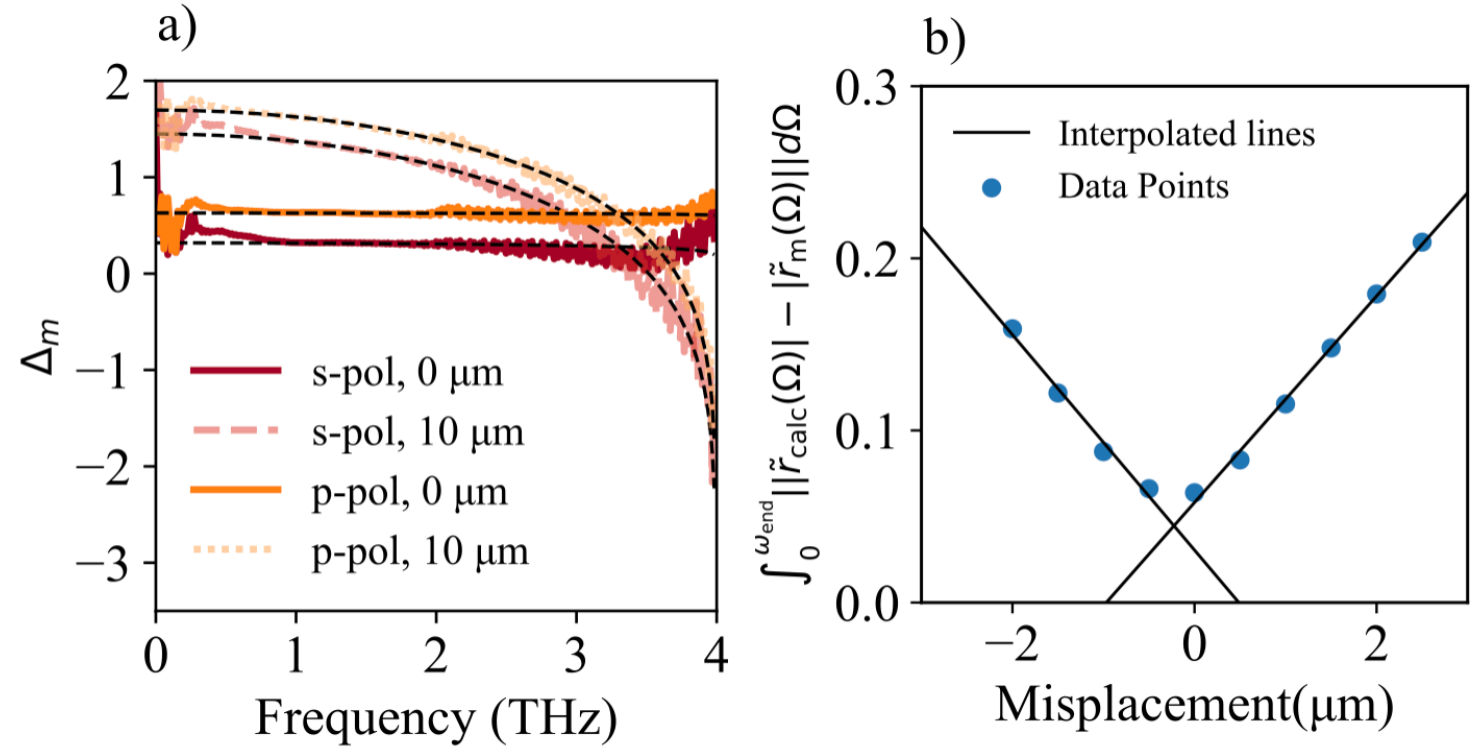}
    \caption{(a) Calculated $\Delta_m$ for InSb in case of $s$- (red) and $p$-polarized (orange) radiation obtained at two positions denoted by solid and dashed lines. Black dashed lines: fit performed with the analytical function in Eq.~(\ref{eq:ikkr_final_simp}).(b) Blue symbols: magnitude of the integrated difference between $|\tilde{r}_\text{calc}|$ and $|\tilde{r}_m|$. Black lines:  linear fits to the symbols. }
    \label{fig:method1M2}
\end{figure}

Equation \eqref{eq:ikkr_final} can be further simplified to use the analytical form as a fitting function for retrieving the value of the misplacement. Due to the finite integration range, the $g(\Omega;\omega, \omega')$  term in the integral in Eq.~\eqref{eq:ikkr_final} can be split into two parts, and all the terms which are not a function of $\omega$ can be collected in a constant denoted by $C'$:
\begin{multline}
    C' = \frac{2}{\pi} \textbf{P} \int_0^{\omega_\text{end}} \frac{\Omega \varphi_m(\Omega)}{\Omega^2-\omega'^2} d \Omega - \ln\left|\tilde{r}_m(\omega')\right| \\ -\frac{2l\omega'}{\pi c \cos\theta} \ln \left|\frac{\omega_{\mathrm{end}}-\omega'}{\omega_{\mathrm{end}}+\omega'}\right|.
\end{multline}
Thus, a simplified version of Eq.~\eqref{eq:ikkr_final} can be written as
\begin{multline}\label{eq:ikkr_final_simp}
    \Delta_m\left(\omega\right) \equiv \frac{2}{\pi} \textbf{P} \int_0^{\omega_\text{end}} \frac{\Omega \varphi_m(\Omega)}{\Omega^2-\omega^2} d \Omega - \ln\left|\tilde{r}_m(\omega)\right| \approx\\
    \frac{2l\omega}{\pi c \cos\theta} \ln \left|\frac{\omega_{\mathrm{end}}-\omega}{\omega_{\mathrm{end}}+\omega}\right| + C + C',
\end{multline}
where we defined $\Delta_m$, the only quantity to be calculated from the measured phase $\varphi_m$ and reflection coefficient $\left|\tilde{r}_m\right|$. Eq. \eqref{eq:ikkr_final_simp} can be exploited to obtain the value of the misplacement by fitting $\Delta_m$ with $l$ and the $\left(C + C'\right)$ constant as free parameters. Figure \ref{fig:method1M2}(a) shows the calculated $\Delta_m$ for two polarizations and two misplacement values, with the results of the fitting procedure overlayed as black dashed lines.

\subsection{Experimental minimization technique}\label{sec:exp_min}
Besides the analytical fitting method, the misplacement can also be retrieved by performing a minimization technique on the data. First, the sample is measured at an arbitrary distance from the reference. Then, the measurement is repeated by scanning the sample along the forward and backward directions, and the complex reflection coefficient is recorded for each position. Since Eq.~\eqref{eq:phi_m} analytically describes the effect of a misplacement, we can also simulate the phase variations at various sample positions without repeating the experiment. Either of these approaches will result in a change in the slope of the phase as a function of the misplacement $l$. Based on Eq.~\eqref{eq:ikkr_calc}, we calculate the reflection coefficient amplitude $|\tilde{r}_\text{calc}|$ for several $l$ values, and we subtract it from the known measured amplitude $|\tilde{r}_m|$. For small shifts, the measured reflection coefficient remains stable, while the calculated reflection coefficient changes significantly. The misplacement that minimizes the difference between the computed and measured reflection coefficient provides the correct phase values. To calculate this minimum, which indicates the correct sample position, we compute
\begin{equation}\label{eq:m1}
   \min_l\left( \int_0^{\omega_\text{end}} ||\tilde{r}_\text{calc}(\Omega)|-|\tilde{r}_m(\Omega)||d\Omega \right).
\end{equation}
The results from Eq.~(\ref{eq:m1}) are plotted in Fig.~\ref{fig:method1M2}(b) as a function of $l$, and the data points are interpolated using a linear fit. The crossing point of the interpolated lines provides an estimate of the shift of the sample. Based on our observations, measuring at least 5 points on either side of the initial position, with at least 0.5 \textmu m spacing between shifts, gives a good estimate of the misplacement.

\section{Results and Discussion}\label{sec:discussion}

\begin{table}[b]
\caption{Misplacement values extracted from the analytical fitting method ($l_1$) and the experimental minimization technique ($l_2$)}\label{tab:table1}
\begin{tabular}{|c|l|r|r|}
\hline
Geometry & \begin{tabular}[c]{@{}c@{}} Sample shift\end{tabular} & \begin{tabular}[c]{@{}c@{}}$l_1/\cos\theta$ (\textmu m)\end{tabular} & \begin{tabular}[c]{@{}c@{}}$l_2/\cos\theta$ (\textmu m)\end{tabular} \\  \hline
\multicolumn{1}{|c|}{\multirow{3}{*}{\begin{tabular}[c]{@{}c@{}}Normal \\ incidence\end{tabular}}} & 0 \textmu m                                                    & 0.29   & 0.22          \\ \cline{2-4} 
\multicolumn{1}{|c|}{}   & 10 \textmu m      & 10.10    & 9.96   \\ \cline{2-4} 
\multicolumn{1}{|c|}{}    & 100 \textmu m    & 100.06   & 99.37   \\ \hline
\multirow{4}{*}{\begin{tabular}[c]{@{}l@{}}45\degree \\ incidence\end{tabular}}      & \begin{tabular}[c]{@{}l@{}}$s$-pol,  0 \textmu m\end{tabular}  & 0.37            &   0.45      \\ \cline{2-4}                                                  & \begin{tabular}[c]{@{}l@{}}$s$-pol, 10 \textmu m\end{tabular} & 11.27 & 11.31 \\ \cline{2-4}
& \begin{tabular}[c]{@{}l@{}}$p$-pol, 0  \textmu m\end{tabular} &  0.08 &  0.03 \\ \cline{2-4} 
& \begin{tabular}[c]{@{}l@{}}$p$-pol, 10 \textmu m\end{tabular} & 10.45 & 10.21 \\ \hline
\end{tabular}
\end{table}

We now discuss and compare the results obtained from the two methods described above. The reflection coefficient for InSb is measured at normal incidence geometry in the companion paper \cite{UDCM}. For the analytical fitting procedure (first method),  the measurements are performed using both $s$ and $p$ polarization at 0 \textmu m, as well as with a 10 \textmu m change of the THz path with respect to the reference position. The measured reflection coefficient is stable for each shift; however, due to the change in the phase, the extracted $|\tilde{r}_\text{calc}|$ is affected, as shown previously in Fig.~\ref{fig:3}. The retrieved $|\tilde{r}_\text{calc}|$  for a change of 10 \textmu m shows a significant deviation from that of 0 \textmu m for both polarizations, especially at higher frequencies. This can also be observed in Fig.~\ref{fig:method1M2}(a), which shows the $\Delta_m$ curves for $s$ and $p$-polarized light at 0 and 10 \textmu m. The black dashed lines, in this case, are obtained by fitting the data using Eq.~(\ref{eq:ikkr_final_simp}), where $l$ and the $\left(C + C'\right)$ constant are used as fitting parameters. The $l_1$ misplacement values found using this method are reported in Table \ref{tab:table1}. The fitted constants are found to be: 0.49, 1.51, 11.02 for the three normal incidence measurements. At 45\degree~incidence, we get 0.31, 1.44 for the two positions with $s$-polarized radiation and 0.62, 1.69 for the two positions with $p$-polarized THz.  

 For the experimental minimization technique (second method), we chose 10 values of $l$, spaced 0.5 \textmu m apart within a range of 5 \textmu m. These shift values are used to calculate the phase, which is then used to extract $|\tilde{r}_\text{calc}|$ for each $l$. The blue data points in Fig.~\ref{fig:method1M2}(b) mark the result of Eq.~(\ref{eq:m1}) for each sample position, and the minimum $l$ appears to be between 0 and -0.5 \textmu m. We interpolate the data points on either side of the minimum to achieve a more precise result. The crossing point of the interpolated lines shown in black is at $l=-0.22$ \textmu m. This procedure is applied for normal incidence as well as 45\degree-incidence geometry for $s$ and $p$ polarization of the THz radiation. The values for $l$ obtained using this technique are shown in the column labeled $l_2$ in Table \ref{tab:table1}. The change in the reflected THz path length depends on the angle of incidence, thus, for easier comparison in the table, we denote $l/\cos\theta$ for both geometries.
The values show good agreement with $l_1$, as well as with the actual change in the THz path. %The shifts estimated from this method are used for correcting the phase and computing the optical properties as done previously.

\begin{figure}[t]
    \centering
    \includegraphics[width=\columnwidth]{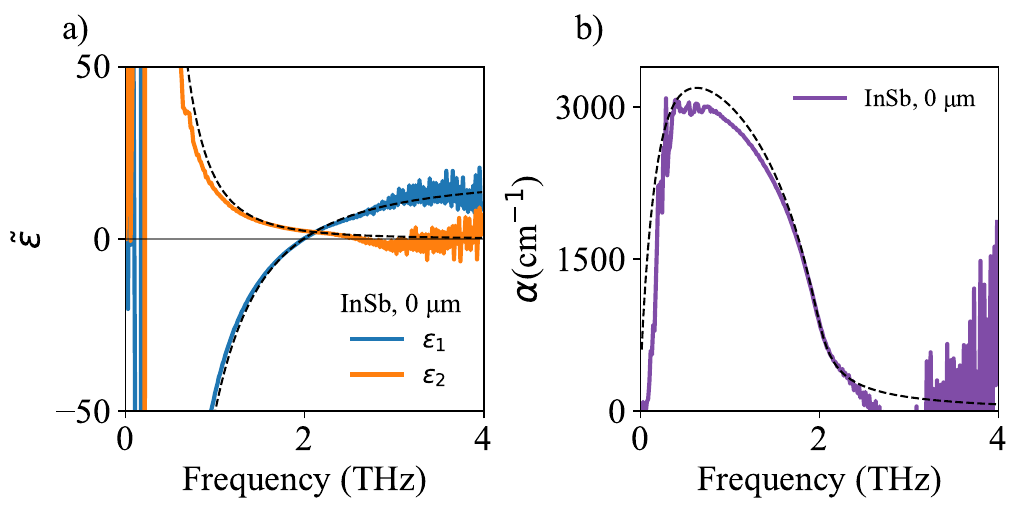}
    \caption{(a) Dielectric function $\tilde{\varepsilon} = \varepsilon_1 + i\varepsilon_2$ and (b) absorption coefficient $\alpha$ for InSb obtained from a normal incidence measurement. The black dashed line denotes the Drude model fitting results.}
    \label{fig:nk_eps_abs}
\end{figure}

The retrieved $l$ values allow to calculate the corrected phase using Eq.~(\ref{eq:phi_m}), which together with the measured amplitude of the reflection coefficient can be used to obtain the complex refractive index. In the case of normal incidence geometry, the complex refractive index $\tilde{n}(\omega)$ is calculated from the reflection coefficient using the Fresnel equations as follows:
\begin{equation}\label{eq:n0}%Eq17
\tilde{n}(\omega)=n(\omega)+ik(\omega)=\frac{1+\tilde{r}(\omega)}{1-\tilde{r}(\omega)}.
\end{equation}
The real ($\varepsilon_1$) and imaginary ($\varepsilon_2$) parts of the dielectric function calculated by taking the square of the complex refractive index are plotted in Fig.~\ref{fig:nk_eps_abs}(a). The curves cross each other at 2.1 THz. The absorption coefficient $\alpha$ shows a broad feature along the low-frequency range, as reported in Fig.~\ref{fig:nk_eps_abs}(b). The black dashed lines in all the panels denote the curves obtained from the Drude model fit, which for the dielectric function is written as
\begin{equation}
    \tilde{\varepsilon}(\omega)=\varepsilon_1(\omega)+i\varepsilon_2(\omega)=\varepsilon_{\infty}\left(1-\frac{\omega^{2}_{p}}{\omega^{2}+i\omega\gamma}\right).
\end{equation}
The fitting parameters are found to be $\varepsilon_\infty=18.16$, $\omega_{p}/2 \pi =2.005$ THz and $\gamma/2 \pi=0.26$ THz. We observe that the optical properties retrieved from the corrected phase show a good overlap both with the fitted curves and with the values reported in Ref.~\cite{houver20192da}.

\begin{figure}[t]
    \centering
    \includegraphics[width=\columnwidth]{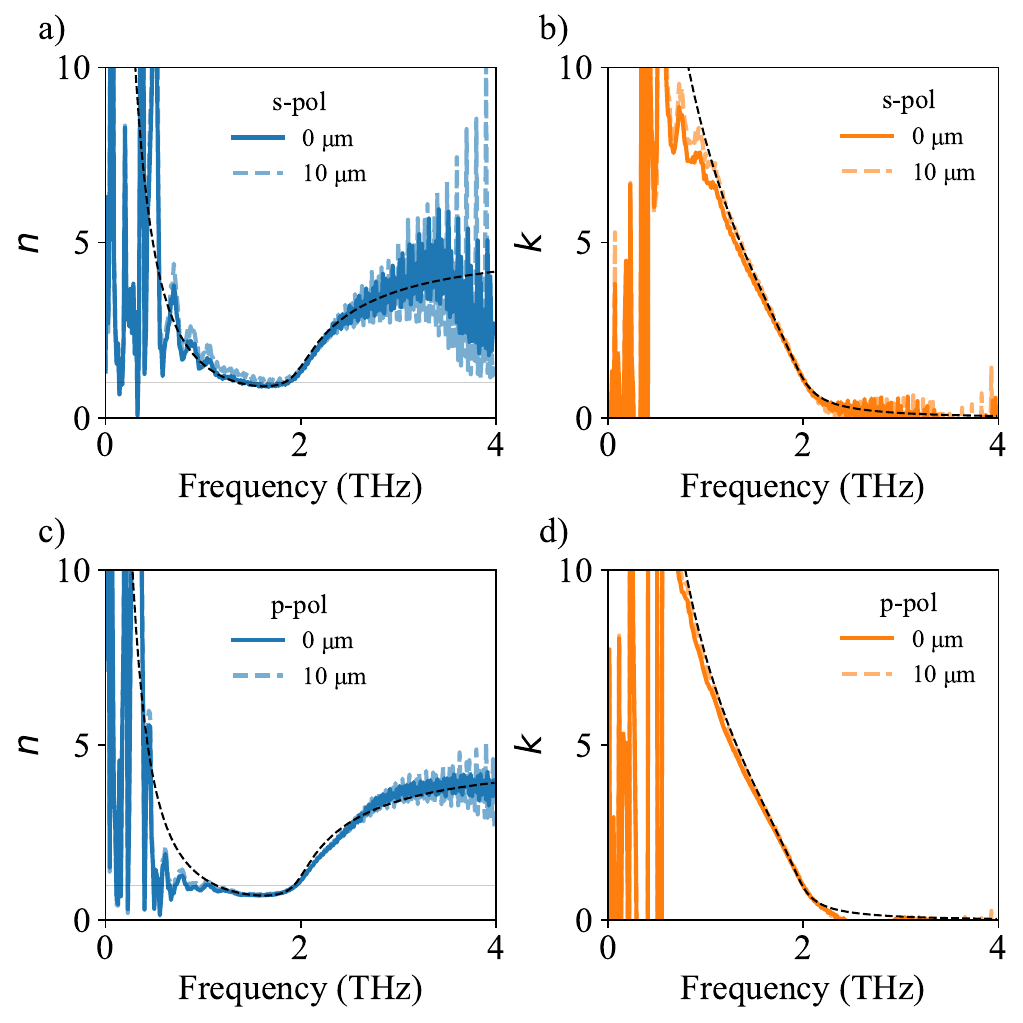}
    \caption{(a, c) Real and (b, d) imaginary part of the complex refractive index of InSb obtained from $s$- and $p$-polarized radiation for two different path lengths: 0 (solid line) and 10 \textmu m (dashed line). The black dashed line denotes the Drude model fitting results.}
    \label{fig:nk_m2}
\end{figure}

For an arbitrary angle of incidence $\theta$, and for $s$- and $p$-polarization states of the radiation, omitting the explicit frequency dependence of $\tilde{n}$ and of the complex reflection coefficients $\tilde{r}_s$ and $\tilde{r}_p$ for the two polarization states, the refractive index can be calculated by
\begin{subequations}
\begin{gather}\label{eq:ns}%Eq16
\tilde{n}_s = \sqrt{1+\frac{4 \tilde{r}_{s} \cos ^{2} \theta}{\left(1-\tilde{r}_{s}\right)^{2}}},\\
\label{eq:np}%Eq17
\tilde{n}_p = \sqrt{\frac{1+\sqrt{1-4 \sin ^{2} \theta \cos ^{2} \theta\left(\dfrac{1-\tilde{r}_{p}}{1+\tilde{r}_{p}}\right)^{2}}}{2 \cos ^{2} \theta\left(\dfrac{1-\tilde{r}_{p}}{1+\tilde{r}_{p}}\right)^{2}}}.
\end{gather}
\end{subequations}

The complex refractive index calculated for the two polarizations is shown in Fig.~\ref{fig:nk_m2}. The black dashed lines denote the results from the Drude model fitting of the measured reflection coefficients for each polarization. The fitting parameters are found to be $\varepsilon_\infty=23.05$, $\omega_{p}/2 \pi =1.98$ THz, $\gamma/2 \pi=0.29$ THz for $s$-polarized radiation, and $\varepsilon_\infty=20.45$, $\omega_{p}/2 \pi=1.98$ THz, $\gamma/2 \pi=0.24$ THz for $p$-polarized one.
The results show a strong absorption feature in each case, particularly close to the plasma frequency at 2 THz. The solid gray line marks $n=1$ for all frequencies, which is marked to highlight the near-zero index behavior of InSb. This also shows that the results match the values obtained for normal incidence measurements as reported in our companion Letter \cite{UDCM}.

The results corroborate the remarkable robustness and accuracy of our approach regardless of incident angle, polarization, and misplacement. Furthermore, it demonstrates the capability of obtaining the misplacement with sub-\textmu m resolution, which is more than two orders of magnitude smaller than the wavelength of the terahertz radiation (1 THz corresponds to 300 \textmu m), and one order of magnitude better than what is expected by the temporal resolution of the setup (50 fs correspond to $\sim$ 15 \textmu m). Although the technique is demonstrated even for large shifts, it works best for a small misplacement. Larger shifts that modify the focus of the optical setup introduce artifacts that require a full three-dimensional modeling to describe the propagation of the radiation.

\section{Conclusion}
We described a systematic and robust phase correction technique, which can be carried out experimentally as well as analytically. We showed the capability of our technique to reliably extract complex refractive indices, dielectric constants, and absorption coefficients for different incident angles, polarizations, shifts, and spectral features. Moreover, our approach could be applied to investigate the influence of misplacement and monitor the experimental surroundings. Here we describe a general technique that accurately determines the phase. However, the results can be further improved based on the knowledge of the specific material being studied. Substituting fitted data, better models, or more complex versions of the Kramers-Kronig relations could lead to further improvement of the results. This further opens up the path for the widespread use of THz-TDS in reflection geometry. We anticipate that our results will provide a substantial contribution to the fundamental understanding of quantum materials, most of which cannot be measured in transmission geometry.
\appendix

\begin{appendices}

\section{}
\subsection*{Comparison of direct and inverse Kramers-Kronig relations}

\begin{figure}
    \centering
    \includegraphics[width=0.95\columnwidth]{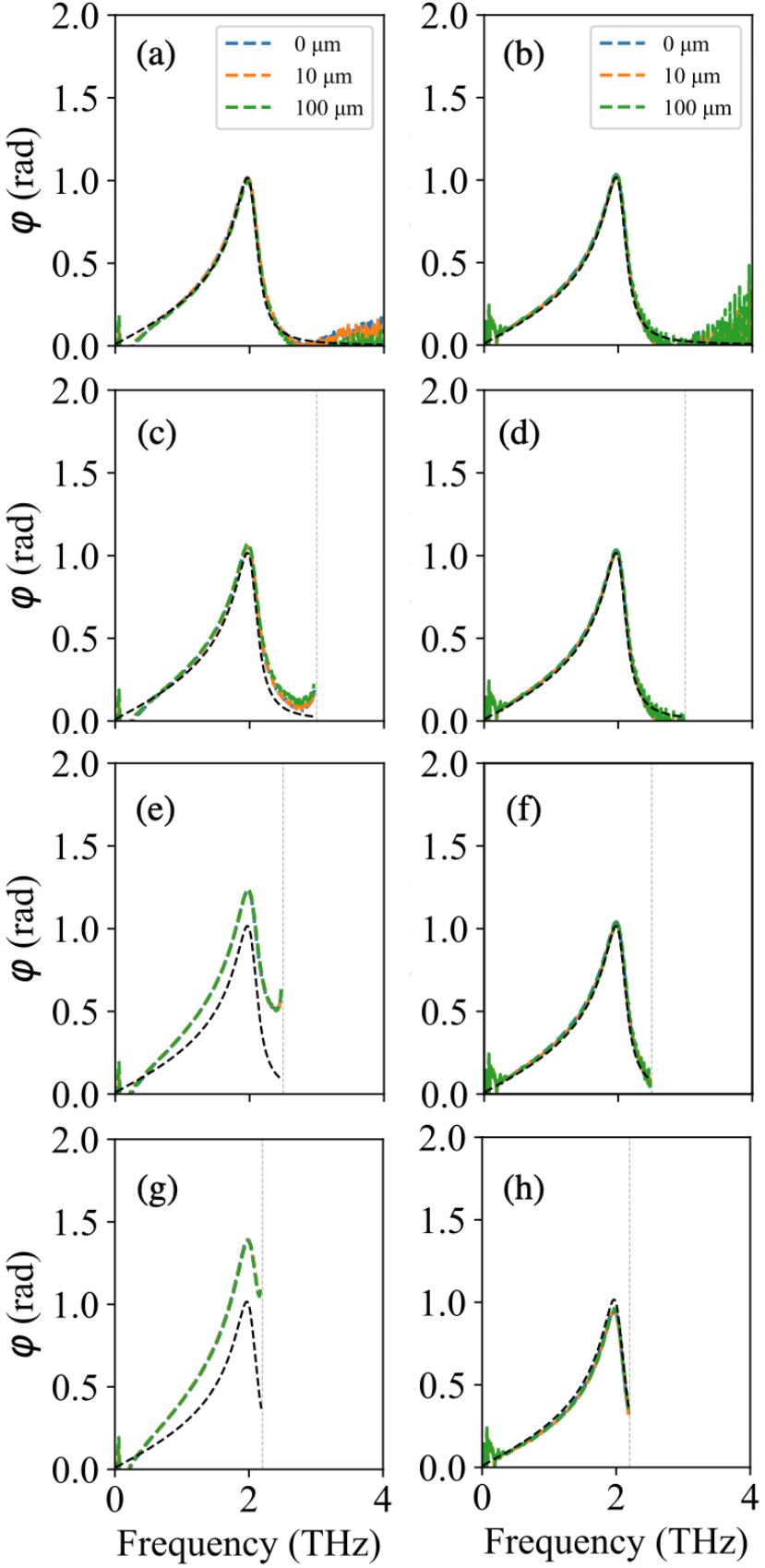}
    \caption{Comparison of the retrieval of the phase using the direct Kramers-Kronig relation (left) and the inverse one, using the experimental minimization technique (right) for sample shifts of 0 \textmu m (blue), 10 \textmu m (orange), and 100 \textmu m (green). The correct value of the phase is shown as a black dashed line. Vertical grey dashed lines indicate truncation frequency $\omega_\text{end}$ at 4 THz (a,b), 3 THz (c,d), 2.5 THz (e,f), and 2.2 THz (g,h).}
    \label{fig:phi_compare}
\end{figure}

In this section we compare the phases obtained using both the direct Kramers-Kronig relation, and the inverse one, using the experimental minimization technique described in Section \ref{sec:exp_min}. The comparison is shown in Fig.~\ref{fig:phi_compare}, where the left panels illustrate the phase extracted from the direct Kramers-Kronig relation, Eq.~\eqref{eq:phi_kkr_full}, while the right panels represent the corrected phase obtained using the experimental minimization technique.
{\label{app:A}}

The retrieved phase is presented in each subplot for measurements in normal incidence geometry, where the sample is shifted by 0, 10, and 100 \textmu m. These are compared to the correct phase shown by the black dashed lines, which is derived from the measured reflection coefficient amplitude via a Drude model. Each row corresponds to a different truncation value for the truncation frequency $\omega_\mathrm{end}$ indicated in the caption. In the left panels, the $\ln|r(\omega_{\mathrm{end}})|$ value needed for the direct Kramers-Kronig equation is obtained by averaging the last 10 points of the measured reflection coefficient, in order to reduce the effect of the noise at the end of the range. While this ensures reasonable phase extraction in cases without spectral features, it can distort the phase shape when spectral features (i.e., the presence of a resonance) are present. Moreover, if we take a closer look at the extracted phases close to the end of the measurement range, the extracted phases in the left panels exhibit a change in shape. In contrast, the phase in the right panels remains reliable almost independently on $\omega_\mathrm{end}$.

\section{Mathematical derivation\label{app:B}}
\emph{Proof of convergence of the integral in Eq.~\eqref{eq:ikkr_lin_inf}}.
We want to show that
\begin{equation}\label{eq:C2}
    I = \textbf{P} \int_0^{\infty} \Omega^2\left(\frac{1}{\Omega^2-\omega^2}-\frac{1}{\Omega^2-\omega^{\prime 2}}\right) d\Omega
\end{equation}
converges to 0. The terms inside the parenthesis can be combined to get
  \begin{equation}\label{eq:C3}
  I = \textbf{P} \int_0^{\infty} \Omega^2\frac{\omega^2-\omega^{\prime 2}}{(\Omega^2-\omega^2)(\Omega^2-\omega^{\prime 2})} d\Omega.
  \end{equation}
We observe that the numerator includes the term $\Omega^2$ while the denominator contains $\Omega^4$, leading to a net effect that suggests the convergence of the integral. Using partial fractions, the integral can be split as
\begin{equation}\label{eq:C4}
    I=\textbf{P} \int_0^{\infty} \frac{\omega^2}{\Omega^2-\omega^2} d\Omega -\textbf{P} \int_0^{\infty}\frac{\omega^{\prime 2}}{\Omega^2-\omega^{\prime 2}} d\Omega.
\end{equation}
Each of these integrals can be further expanded by using partial fractions. Since they have the same form, we show the detailed formulation for one integral only
\begin{equation}\label{eq:C5}
      \frac{\omega^{2}}{\Omega^2-\omega^{2}}
   = \frac{\omega}{2} \frac{1}{(\Omega-\omega)} - \frac{\omega}{2} \frac{1}{(\Omega+\omega)}.
\end{equation}

By rewriting the first integral in Eq.~\eqref{eq:C4} as the two terms in Eq.~\eqref{eq:C5}, we can simplify the integration to obtain
\begin{equation}\label{eq:C6}
    \frac{\omega}{2}\left[ \textbf{P}\int_0^{\infty} \frac{d\Omega}{\Omega-\omega}- \textbf{P}\int_0^{\infty} \frac{d\Omega}{\Omega+\omega}\right] = \frac{\omega}{2}\ln\left|\frac{\Omega-\omega}{\Omega+\omega} \right|_0^{\infty}=0.
\end{equation}
The same procedure can be followed for the second integral in Eq.~\eqref{eq:C4}, and we thus conclude that the integral $I$ converges to 0 when the limits of integration are from 0 to $\infty$.

\emph{Derivation of Eq.~\eqref{eq:ikkr_final}}.
We want to solve the integral within Eq.~\eqref{eq:ikkr_grouped} in the main text
\begin{equation}\label{eq:D1}
I = \textbf{P} \int_0^{\omega_\text{end}} \Omega^2\left(\frac{1}{\Omega^2-\omega^2}-\frac{1}{\Omega^2-\omega^{\prime 2}}\right) d\Omega
\end{equation}
in a finite frequency range, which allows us to split the integral into two separate terms:
\begin{equation}\label{eq:D2}
I=\textbf{P} \int_0^{\omega_\text{end}} \frac{\Omega^2}{\Omega^2-\omega^2} d\Omega  -\textbf{P}\int_0^{\omega_\text{end}} \frac{\Omega^2}{\Omega^2-\omega^{\prime 2}}d\Omega.
\end{equation}
Since the two terms are of the same form, we demonstrate the solution for one of them. We proceed in the following way:
\begin{multline} \label{eq:D3}
    \mathbf{P} \int_0^{\omega_\text{end}} \frac{\Omega^2}{\Omega^2-\omega^{2}}d\Omega = \mathbf{P} \int_0^{\omega_\text{end}} \frac{\Omega^2 - \omega^{2}+\omega^{2}}{\Omega^2-\omega^{2}}d\Omega =\\
    =\int_0^{\omega_\text{end}} 1 d\Omega+ \mathbf{P} \int_0^{\omega_\text{end}} \frac{\omega^{2}}{\Omega^2-\omega^{2}}d\Omega.
\end{multline}
Using partial fractions as described in Eq.~\eqref{eq:C5}, we can expand the second term in the integral. Combining all the terms we get:
\begin{multline}
    \mathbf{P} \int_0^{\omega_\text{end}} \frac{\Omega^2}{\Omega^2-\omega^{2}}d\Omega =\\ =\int_0^{\omega_\text{end}} 1 d\Omega + \frac{\omega}{2}\mathbf{P}\int_0^{\omega_\text{end}} \frac{d\Omega}{\Omega-\omega} - \frac{\omega}{2}\mathbf{P}\int_0^{\omega_\text{end}} \frac{d\Omega}{\Omega+\omega}.
\end{multline}
Finally, we obtain 
\begin{equation}
  \mathbf{P} \int_0^{\omega_\text{end}} \frac{\Omega^2}{\Omega^2-\omega^{2}}d\Omega = \omega_\text{end} + \frac{\omega}{2}\ln\left|\frac{\omega_\text{end}-\omega}{\omega_\text{end}+\omega}\right|.
\end{equation}
Similarly, for the second integral we get 
\begin{equation}
  \mathbf{P} \int_0^{\omega_\text{end}} \frac{\Omega^2}{\Omega^2-\omega^{\prime 2}}d\Omega = \omega_\text{end} + \frac{\omega^{\prime}}{2}\ln\left|\frac{\omega_\text{end}-\omega^{\prime}}{\omega_\text{end}+\omega^{\prime}}\right|.
\end{equation}
Subtracting both the terms gives the result we are looking for:
\begin{equation}
    I=  \frac{\omega}{2}\ln\left|\frac{\omega_\text{end}-\omega}{\omega_\text{end}+\omega}\right| -  \frac{\omega^{\prime}}{2}\ln\left|\frac{\omega_\text{end}-\omega^{\prime}}{\omega_\text{end}+\omega^{\prime}}\right|.
\end{equation}
\end{appendices}

\providecommand{\noopsort}[1]{}\providecommand{\singleletter}[1]{#1}%

\end{document}